\documentclass[journal]{IEEEtran}

\usepackage{graphicx,psfrag,epsf}

 \usepackage{setspace}
\usepackage{url} 
\usepackage{amsmath,textcomp,amssymb,graphicx,enumerate,mathtools,bm,multirow, caption}
\usepackage{algorithm} 
\usepackage{algcompatible}
\usepackage[margin=1in]{geometry}
\renewcommand{\algorithmicensure}{ \textbf{Setting:}} 
\usepackage{hyperref}
\usepackage{color}
\usepackage{mathrsfs}
\usepackage{booktabs}
\usepackage{chngcntr}
\usepackage{pbox}

\usepackage[normalem]{ulem}
\hypersetup{
	colorlinks=true,
	linkcolor=blue,
	filecolor=magenta,      
	urlcolor=blue,
	citecolor=blue
}
\usepackage{bbm}
\usepackage{subfig}
\usepackage{supertabular}
\usepackage{tabularx}
\DeclareCaptionFont{xipt}{\fontsize{10}{12}\mdseries}

\begin{document}
\def\spacingset#1{\renewcommand{\baselinestretch}%
{#1}\small\normalsize} \spacingset{1}


	\title{Conditional Kernel Density Estimation Considering Autocorrelation for Renewable Energy Probabilistic Modeling}
	\author{Yuchen~SHI,~\IEEEmembership{Student Member,~IEEE,}, and Nan~CHEN,~\IEEEmembership{Member,~IEEE,}}
	\date{}
	\maketitle

\begin{abstract} 
	Renewable energy is essential for energy security and global warming mitigation. However, renewable power generation is uncertain due to volatile weather conditions and complex equipment operations. It is therefore important to understand and characterize the uncertainty in renewable power generation to improve operational efficiency. In this paper, we proposed a novel conditional density estimation method to model the distribution of power generation under various weather conditions. It explicitly accounted for the temporal dependence in the data stream and used an iterative procedure to reduce the bias in conventional density estimation. Compared with existing literature, our approach is especially useful for the purpose of short-term modeling, where the temporal dependence plays a more significant role. We demonstrate our method and compare it with alternatives through real applications.
\end{abstract}
\noindent%
\singlespacing
\section{Introduction}\label{sec:intro}
Renewable energy (RE) plays an essential role in enhancing energy
security and mitigating global warming. By 2050, renewable energy
sources (RES) and electrification is expected to provide 75\% of the
necessary reductions in energy-related carbon emissions to limit the
global rise in temperature \cite{IRENA}. Nevertheless, one major
challenge of the increasing RE penetration in the energy portfolio is the
highly uncertain power generation, which raises challenges for system
balancing of supply and demand. This is especially true for renewable
energy that can not be stored after power generation, such as wind and solar energy. Other examples include tidal energy and wave energy.

Therefore, understanding and characterizing uncertainty in RES is an
essential topic. Generally speaking, the uncertainties contained in
typical RES are two-fold. Firstly, the weather conditions are
uncertain. RES are usually harvested from natural processes, such as
sunlight, wind, rain, tides, waves, and geothermal heat. Due to this
nature, forecasting weather conditions plays a crucial rule in RE
forecasting, yet they are not the focus of this work. Interested
readers can refer to \cite{lei2009review,
  zhang2017critical} and references
therein. Secondly, the functional relationships between power
production and various weather conditions are unclear. In other words,
even if accurate weather conditions can be obtained, the amount of
power generated is still uncertain. We adopt the term ``power
  curve" to refer to this functional relationship in this work
\cite{lee2015power}. With some generalization, in this work, we define
``power curve" as the probabilistic relationship between renewable
energy production and different weather conditions. Estimating power
curve for uncertainty quantification is the main focus of this
work. The power curve estimation is important and indispensable in
several impact areas for RES. Firstly, should power curve be combined
with prediction of weather conditions, it can be used for RE
forecasting, which helps to improve the control of RE
systems, increase the security and efficiency of power grid, and
benefit power pricing and trading
\cite{Hong2016,lee2018bivariate}. Moreover, power curve serves as a
system performance indicator which is critical for condition monitoring, reliability and maintenance \cite{sohoni2016critical,yildirim2017integrated}.

\begin{figure}[!t]   
	\centering  
\subfloat[Heat plot]{\includegraphics[width=0.48\linewidth]{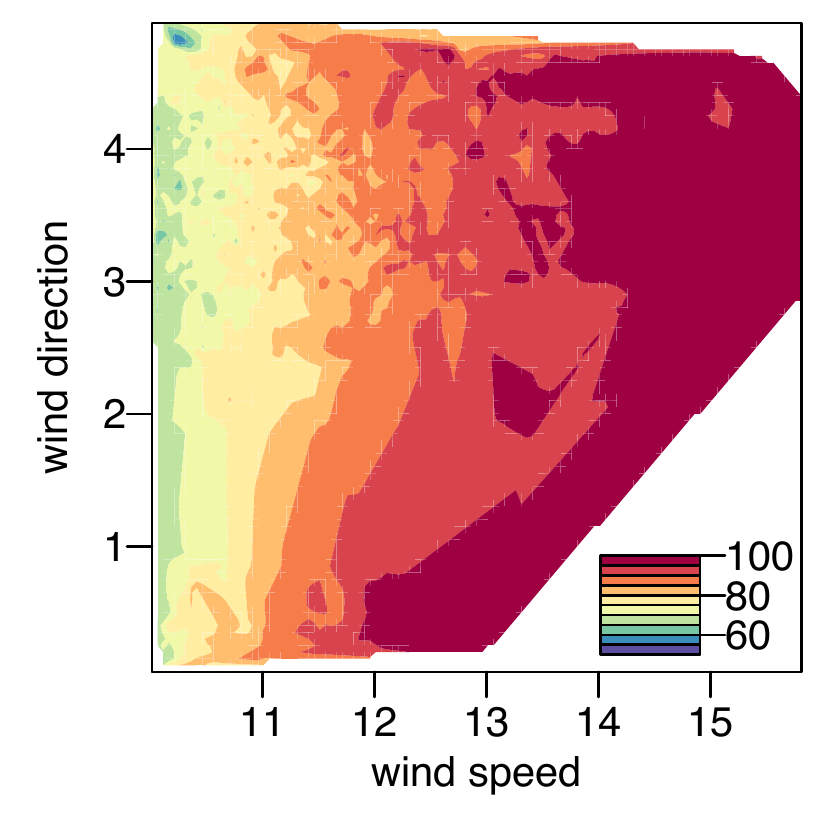}
	\label{Wind_a}}
\subfloat[Density plot conditional on different wind speed (V)]{\includegraphics[width=0.48\linewidth]{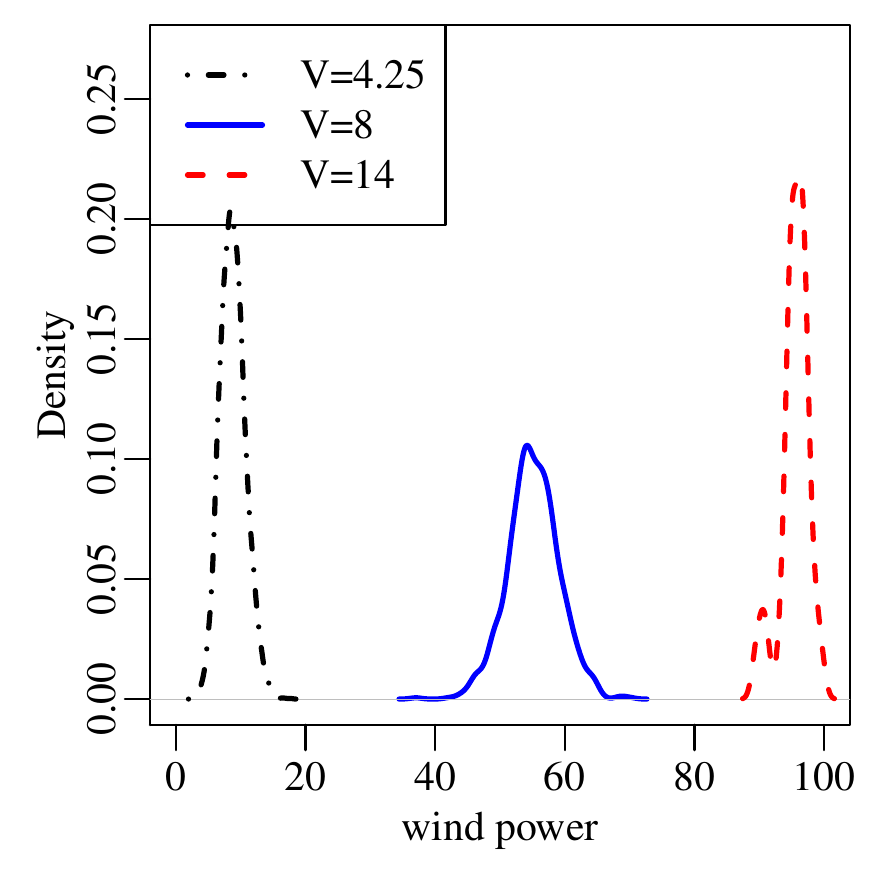}\label{Wind_b}}
	\caption{Visualization of wind power production }\label{Wind}
\end{figure}

Estimating the power curve is inherently challenging because the mechanisms to harvest RE are usually complex. Taking wind power as an example, power generation of a wind turbine is a function of several variables: radius of the rotor, air density and wind
speed. Nevertheless, their relation does not follow a known functional form and depends ambiguously on turbine type, blade pitch, tip-speed ratio, etc \cite{ackermann2005wind}. Therefore, modeling the power curve based on first principles is inaccurate. Researchers have also adopted statistical regression and data-driven methods for power curve modeling. Generally, modeling power curve has three challenges: (1) The system inputs, i.e., weather conditions, are highly interactive. As an example, the interaction between wind speed and wind direction on wind power production is shown in Fig. \ref{Wind_a}. (2) The conditional distributions of power production under different weather condition are heterogeneous. Consider the power production under different  weather conditions as conditional random variables, their conditional distributions typically have different means, variances, even
different distributional forms, as shown in the Fig. \ref{Wind_b}. These two challenges make the conventional parametric methods less appealing. Instead, nonparametric conditional density estimation methods can better deal with ambiguous interactions without distributional assumptions, and even work for the heterogeneous conditional distributions. With these considerations, we use conditional density estimation for power curve estimation in this work. (3) Even if the temporal dependence in weather variables are accounted, short-term temporal dependence in power production still exists, which is possibly caused by equipment's inertia. Without properly accounting for such dependence, the inference could be less accurate or even misleading.

The first and second challenges have been studied in the
literature. Overall, literature on power curve estimation can be
classified into two categories. Methods in the first category focus on the conditional mean of power production ($Y$) at given
weather condition ($\bm{X}$), i.e., $\mathbb{E}(Y|\bm{X})$. Within this category,  physical models and statistical models have been widely used. Interested readers can refer to \cite{Lydia2014comprehensive, giebel2011state} for nice reviews. Besides, the recent trend is to use the machine learning and deep learning approaches. \cite{ferlito2017comparative} compared 11 machine learning methods and concluded that in the online mode with proper window length, SVM is the best performing model, followed by the cubist. \cite{shamshirband2019survey} and \cite{wang2019review} did nice reviews for the deep learning methods and pointed out the main streams of the most typical applied techniques, i.e., convolutional neural networks, recurrent neural networks, deep belief neural networks, and autoencoders. Should the conditional mean estimation be combined with parametric assumption, the conditional variance and confidence interval can also be constructed \cite{wan2013probabilistic, bae2016hourly, wang2017deep}

Nevertheless, methods in the first category fails to capture the heterogeneous conditional distribution of power
output. To this end, methods in the second category provide nonparametric conditional density $f(y|\bm{X})$ at input
($\bm{X}$). These methods require minimal distributional assumptions and manage to make flexible distributional inference. Within in this category, kernel based methods are state-of-art methods used to provide a smooth density function in one step. Due to their flexibility and efficiency, kernel density estimation methods have been widely used and proved to be effective for renewable energy modeling \cite{jeon2012using, pinson2012probabilistic, bessa2012time, lee2015power, Jeon2016, khorramdel2018fuzzy,wahbah2019root}. Moreover, although without providing the full distribution, quantile regression is another popular type of method that focus on one or a few quantile points. It can also be combined with typical machine learning methods by substituting the objective function with the quantile loss, although estimating many quantiles at the same time may increase the challenge for optimization and resulting in cross quantiles \cite{wan2013optimal, zhang2014advanced, golestaneh2016very, toubeau2018deep}. In this work, we focus on the kernel based type of method, which is the state-of-art to provide comprehensive information of the density for the decision makers.

Most of the conditional kernel density estimation literature assumed that the power generations are conditional independent given the meteorological variables. It is well understood that the meteorological variables are temporally dependent. Therefore, the power generation from the temporally dependent weather conditions are naturally dependent. However, it is often overlooked that even if the temporal dependence in weather variables are fully accounted, the conditional
power generations are still dependent due to inertia in hardware or
software of RE systems. To differentiate, we can call the first source
of dependence as ``input dependence'', while the second source due to
system inertia as ``system dependence''. In a short time period, the
system dependence cannot be neglected. To the best of our knowledge, no work considers such dependence in the
conditional density estimation. To fill in the gap in the existing literature, in this paper, we propose a novel nonparametric conditional density estimation method for RE, which explicitly accounts for the short-term system dependence among conditional distributions. We first account for the temporal dependence in the conditional mean, and subsequently establish the conditional density. It is worth noticing that with an emphasize on time dependency, the current work focuses on single devices. Incorporating spatial correlations is certainty important and non-trivial, which will be left for future investigation. The remainder of this paper is organized as follows. In Section 2, we review the conventional kernel method. In Section 3, we introduce our proposed method. In Section 4, we demonstrate the effectiveness of our proposed method by case studies. In Section 5, we conclude the paper with discussions. 
\section{A Review of Conventional Kernel Method}\label{method}
\begin{figure}
	\centering
	\subfloat[Constant mean scenario]{\includegraphics[width=0.48\linewidth]{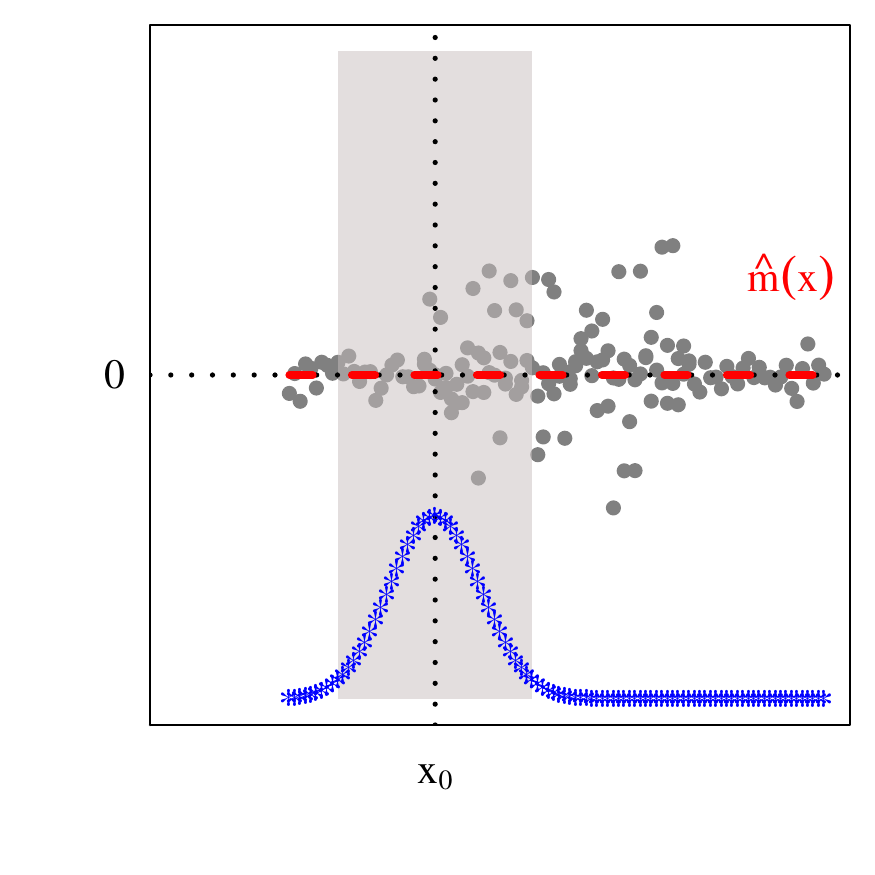}\label{bias_a}}
	\subfloat[Asymmetry effects scenario]{\includegraphics[width=0.48\linewidth]{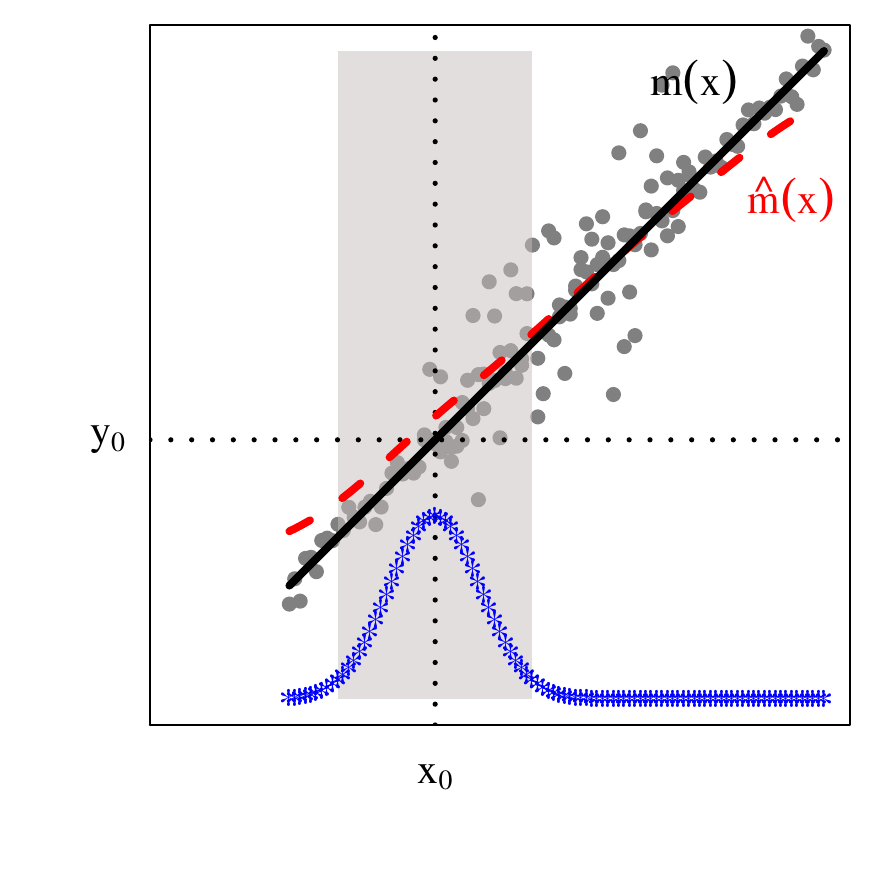}\label{bias_b}}
	\caption{\small{The bias of conventional kernel method. Solid
			(black) curve: real function; Dashed (red) curve: conventional kernel estimation. Dotted (blue) curve: the weights assigned by Gaussian kernel.} }\label{bias}
\end{figure}

\begin{figure}[t]
	\centering
	\subfloat[The ACF plot]{\includegraphics[width=0.48\linewidth]{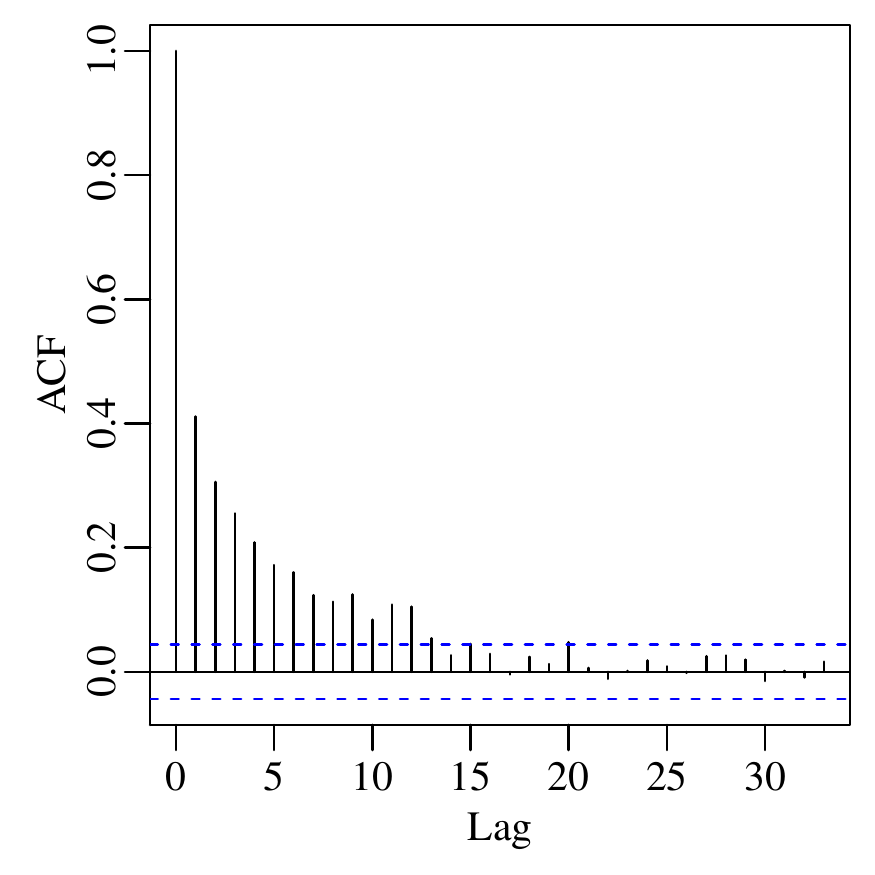}}
	\subfloat[Consecutive residuals]{\includegraphics[width=0.48\linewidth]{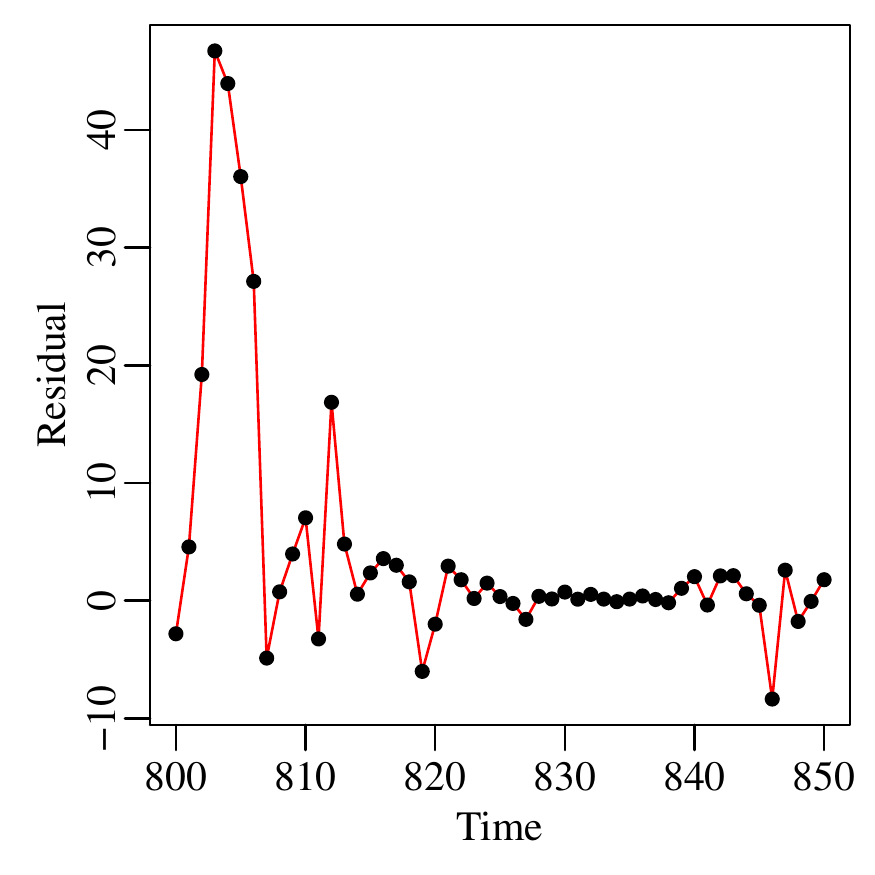}}
	\caption{\small{Temporal dependence in wind power data}}\label{ACF}
\end{figure}

The idea of kernel density estimation can be traced back to \cite{rosenblatt1969conditional}, and has been widely used for renewable energy modeling recently \cite{jeon2012using, pinson2012probabilistic, Jeon2016}. Suppose we have $N$ observations $\{\bm{X}_i,Y_i\}$, $i=1\cdots N$, where $\bm{X}_i\in \mathbb{R}^d$ is
a vector of input variables and $Y_i\in \mathbb{R}$ denotes the
corresponding output value. Let $\hat f_c$ denote the conventional conditional kernel density estimator:
\begin{equation} \label{St_ker}
\hat{f}_c (y|\bm{X})=\sum_{i=1}^{N} w_i(\bm{X})\mathcal{K}(y-Y_i;h_y),
\end{equation}
where
\begin{equation} \label{w}
w_i(\bm{X})=\frac{\mathcal{K}(\|\bm{X}-\bm{X}_i\|;\bm{h_X})}{\sum_{i=1}^{N}\mathcal{K}(\|\bm{X}-\bm{X}_i\|;\bm{h_X})}.
\end{equation}
Here, $\mathcal{K}(\cdot;h)$ is the kernel function that is assumed to be a real-valued,
integrable and non-negative symmetric function. $\bm{h_X}$ and $h_y$ denote the smoothing parameter (bandwidth) to be chosen adaptively. They
are crucial parameters that control the performance of estimation. A larger bandwidth makes the kernel estimator smoother, but increases the bias. A smaller bandwidth does the opposite. Based on (\ref{St_ker}), we can obtain the conditional mean estimator by integration:
\begin{equation} \label{St_Reg}
\hat{\mathbb{E}}_ c(Y|\bm{X})=\int y\hat{f}(y|\bm{X})dy=\sum_{i=1}^{N} w_i(\bm{X})Y_i,
\end{equation}
Despite its simplicity, conventional kernel estimators have some
limitations. Firstly, when the conditional mean is not constant, conventional kernel estimation can be biased because of possible asymmetry effect, curvature effect, or boundary effect in the data \cite{hastie1993local}. Fig. \ref{bias} shows an example of bias caused by asymmetry effect. There are more samples on the right-hand side of $x_0$ than left. When the conditional mean is constant, there is no estimation bias, as in Fig. \ref{bias_a}. In contrast, when the conditional mean is not constant, the asymmetric samples cause a biased  estimation, i.e., $\hat{\mathbb{E}}(y|x_0)>y_0$, as in Figure \ref{bias_b}. Consequently, conditional density estimation would  also be biased. Besides the  asymmetry effect, bias can be  introduced when $x_0$ is near domain boundary or when the mean function has severe curvature. Being aware of this limitation, various literature aims to reduce the bias, such as \cite{khorramdel2018fuzzy,wahbah2019root}.
In addition, conventional kernel method suffers from ``curse of dimensionality'', even if the input dimension of $\bm{X}$ is as high as 4 or
5. Given the definition in \eqref{w}, $w_i$ should asymptotically place dominant mass in a convex region centered
at $\bm{X}$, controlled by the smoothing parameter $\bm{h_X}$. As the
dimension $d$ increases, the probability of having enough sample in a
unit area in the $d$-dimensional space diminishes exponentially. There
were some attempts to tackle this difficulty. For example, \cite{lee2015power} introduced an additive structure on the
conventional kernel methods, where each kernel is restricted to have at most three dimensional inputs. Last but not least, the conventional kernel estimation assumes data are independent and ignores temporal dependence. In RE data streams, autocorrelation among data is common, as is shown in Fig. \ref{ACF}. Here the conditional mean is estimated using Additive Multiplicative Kernel (AMK) \cite{lee2015power}, which is the state-of-art method for wind power curve modeling in the literature. Fig. \ref{ACF} shows the autocorrelation function (ACF) of residuals and a segment of consecutive residuals. Both indicate possibly strong temporal dependence. To the best of our knowledge, the existing literature did not well address this problem, thus we aim to fill this gap to account for the autocorrelation, and also keep the first two limitations into consideration.

\section{Conditional Kernel Density Estimation Considering Autocorrelation (DEAR)}
\subsection{The proposed model}
Based on the review of conventional kernel method, our new method aims to explicitly account for the data autocorrelation in conditional density estimation. At the same time, it provides an flexible structure to reduce the the bias from non-constant mean, and an additive structure proposed by \cite{lee2015power} when necessary to avoid ``curse of dimensionality". Henceafter, we name the proposed method as \textit{conditional kernel Density Estimation considering AutocoRrelation} (DEAR). Let
$\mathcal{F}_T$ denote the set that contains all historical data up to time $T$, $\mathcal{F}_T:=\{(\bm{X}_t,Y_t), t=1,\cdots,T\}$, where
$\mathbf{X}_t\in \mathbb{R}^d$ and $Y_t\in \mathbb{R}$. Assume the data follows the model:
\begin{equation} \label{model}
Y_{t}=m(\bm{X}_{t})+\sigma(\bm{X}_{t})u_{t},
\end{equation}
where $m(\cdot)$ and $\sigma(\cdot)$ are unknown smooth functions to
represent conditional mean and conditional standard deviation, respectively. $u_t$ is a stationary
process with mean 0. $u_{t}$ can be expressed in the autoregressive representation with order $p$:
\begin{equation} \label{model_r}
u_{t}=\sum_{\tau=1}^{p} a_\tau u_{t-\tau}+\epsilon_{t},
\end{equation}
where $\epsilon_t$ are independently and identically distributed
(\textit{i.i.d}) with mean zero and follow unknown distribution. $p$
is a fixed constant depending on autocorrelation strength of the data.
Following this model, the conditional mean can be expressed as    
\begin{align}   \label{model_m}       
\begin{split}                   
&\mu_{t|t-1}:=\mathbb{E}(Y_t|\bm{X}_t, \mathcal{F}_{t-1})\\
&=m(\bm{X}_t)+\sigma(\bm{X}_{t})\sum_{\tau=1}^p \frac{a_\tau[Y_{t-\tau}-m(\bm{X}_{t-\tau})]}{\sigma(\bm{X}_{t-\tau})}.
\end{split}
\end{align}      
Given $\mu_{t|t-1}$, define $F_{t|t-1}(y):=F(y|\bm{X}_t, \mathcal{F}_{t-1})$ as the conditional cumulative distribution function (CDF) of $Y_t$:
\begin{align}\label{model_cdf}
\begin{split}
  F_{t|t-1}(y)=
  G\left(\frac{y-\mu_{t|t-1}}{\sigma(\bm{X}_t)}\right),
\end{split}
\end{align}
where $G$ is the CDF of $\epsilon_t$. Taking derivatives with respect to
$y$ on both sides, we have the conditional density 
\begin{align}\label{model_e}
f_{t|t-1}(y)=g\left(\frac{y-\mu_{t|t-1}}{\sigma(\bm{X}_{t})}\right)/ \sigma(\bm{X}_{t}).
\end{align}
Here $f_{t|t-1}(y):=f(y|\bm{X}_t, \mathcal{F}_{t-1})$ stands for the
conditional density of $Y$ and $g$ stands for the probability density
function of $\epsilon_t$. 
\eqref{model_e} shows that estimating $f_{t|t-1}(y)$ directly links to estimating $g(\epsilon_t)$. \eqref{model_m}-\eqref{model_e} shows the essential difference between DEAR and conventional kernel method. For DEAR, an autocorrelation process is firstly imposed on conditional mean. Conditional density is then built on the residual of conditional mean, which is expected to be uncorrelated. 
\subsection{Inference Procedure of DEAR}
 DEAR provides the flexibility to apply higher order local regression
 for estimating $m(\cdot)$, which reduces the bias when the
 conditional mean is not a constant. In our case studies,
 $m(\cdot)$ is estimated by local linear regression. From our experience, local linear regression can effectively reduce the bias and is sufficient for power curve estimation. Higher order local polynomial regressions cause more severe ``curse of dimensionality". For notation simplicity, we define
$$\bm{\mathcal{Y}}=\left(\begin{matrix} Y_1 \\ \vdots\\Y_T\end{matrix}\right),\bm{\mathcal{X}}=\left(\begin{matrix} 1\ \ (\bm{X}-\bm{X}_1)' \\ \vdots\\1\ \ (\bm{X}-\bm{X}_T)'\end{matrix}\right),$$ and $\bm{\mathcal{W}}$ as a $n\times n$ diagonal
matrix with the $i$th diagonal element being $\mathcal{K}(\|\bm{X}-\bm{X}_1\|;\bm{h_X})$. When the data are independent, an estimator of $m(\cdot)$ obtained by local linear regression can be written as:
\begin{equation} \label{ll}
\tilde{m}(\bm{X})=\bm{e}_1'(\bm{\mathcal{X}}'\bm{\mathcal{W}}\bm{\mathcal{X}})^{-1}\bm{\mathcal{X}}'\bm{\mathcal{WY}}, 
\end{equation}
where $\bm{e}_1$ is a $(d+1)\times1$ vector with 1 in the first element and zeros elsewhere. In the following, we represent the conditional mean estimator as $\tilde{m}(\cdot)$ when estimated using dataset $\mathcal{F}_T$. Similarly, $\sigma^2(\cdot)$ can be estimated by the squared residuals:
\begin{equation}\label{m}
{Z}_t:= [Y_t-\tilde{m}(\bm{X}_{t})]^2\ \ \ \textnormal{for } t=1, \cdots, T.
\end{equation}
The conditional variance estimator estimated by $\mathcal{G}_T:=\{(\bm{X}_t,Z_t), t=1,\cdots,T\}$ is then denoted as $\tilde{\sigma}^2(\cdot)$.

$\tilde{m}(\cdot)$ and $\tilde{\sigma}(\cdot)$ are standard estimators when data are
independent, yet less efficient when data are temporally dependent. To account for the autocorrelation, an iterative procedure is adopted to estimate the $m(\cdot)$,
$\sigma(\cdot)$, and the AR parameters $a_\tau$.
The iterative procedure first uses estimates of  ${m}(\cdot)$ and
${\sigma}(\cdot)$ to update the estimates of the AR parameters
$a_\tau$, then the updated AR parameters in return can be used to  update the estimation of $m(\cdot)$ and $\sigma(\cdot)$. Taking $\tilde{m}(\cdot)$ and $\tilde{\sigma}(\cdot)$ as the initial estimates of $m(\cdot)$ and $\sigma(\cdot)$, the two-step iterative
procedure is summarized as follows. 

\noindent\textbf{Step 1: Estimate AR parameters given $\hat m(\cdot)$ and $\hat\sigma(\cdot)$}

Given existing estimates $\hat m(\bm{X})$ and $\hat \sigma(\bm{X})$
of the mean function and standard deviation function, 
we can obtain $\hat{u}_t=\hat{\sigma}(\bm{X}_{t})^{-1}[Y_t-\hat{m}(\bm{X}_{t})]$ for
$t=1,\cdots, T$ by model \eqref{model}-\eqref{model_r}. $\hat u_t$ is stationary  when $\hat
m(\bm{X})$ and $\hat \sigma(\bm{X})$ are consistently estimated. 
To allow flexible distribution of $\epsilon_t$, we
choose the least square approach:
\begin{equation}
  \label{eq:ls}
  \hat{a}_1,\cdots\hat{a}_p=\arg\min_{{a}_1,\cdots {a}_p}
  \sum_{t=p+1}^T\left[\hat u_t-\cdots {a}_{p}\hat{u}_{t-p}\right]^2
\end{equation}

\noindent\textbf{Step 2: Update $\hat m(\cdot)$ and $\hat \sigma(\cdot)$ given
AR parameters}

Given the estimated AR parameters in Step 1 and previous estimates
$\hat m(\bm{X})$ and $\hat \sigma(\bm{X})$, we are able to improve the
estimation of the mean function and standard deviation function. 
According to (\ref{model}), $m(\bm{X}_t) =
Y_t-\sigma(\bm{X}_{t})\sum_{\tau=1}^{p}{a}_\tau{u}_{t-\tau}$. We
define 
\begin{equation}
\tilde{Y}_t=Y_t-\hat\sigma(\bm{X}_{t})\sum_{\tau=1}^{p}\hat{a}_\tau\hat u_{t-\tau}.
\end{equation}
Because of the autocorrelation, it can be shown that $\mathbb E(\tilde
Y_t)$ is closer to $m(\bm{X}_t)$ than $\mathbb E(Y_t)$. Substituting $Y_t$ by $\tilde Y_t$ thus yields a more
accurate estimator. Likewise, $\hat \sigma(\bm X)$ can be updated
using $\tilde Z_t= [\tilde Y_t-\hat{m}(\bm{X}_{t})]^2$.

The iteration continues until convergence.
At the same time, we do Box-Ljung test on all the $p$ lags
of the residual  series $r_t:= (Y_t-\hat{\mu}_{t|t-1})/\hat{\sigma}(\bm{X}_t)$. The iteration terminates when the AR
parameters converge and the Box-Ljung
tests for autocorrelation at all $p$ lags are not rejected.

From our experience, a  few iterations are sufficient to improve performance with finite samples in practice. It has been proved that under mild conditions, $\hat{m}(\bm{X}_t)$
estimated by this iterative procedure has the same asymptotic bias as
the estimator $\tilde{m}(\bm{X}_t)$ for \textit{i.i.d} data. But $\hat{m}(\bm{X}_t)$ has smaller variance
than $\tilde{m}(\bm{X}_t)$ because autocorrelations are taken into
account, which makes it more efficient \cite{xiao2003more}. 
After the iterations terminate, we can obtain the conditional mean estimator as
\begin{equation} \label{est_mu}
\hat{\mu}_{t|t-1}=\hat{m}(\bm{X}_t)+\hat{\sigma}(\bm{X}_{t})\sum_{\tau=1}^p \hat{a}_\tau\hat{u}_{t-\tau}.
\end{equation}
To estimate $f_{t|t-1}(y)$, it is sufficient to estimate
$g(\epsilon_t)$, with a scale transformation. Given the \textit{i.i.d}
assumption of $\epsilon_t$, conventional kernel estimator can be
used. Using the homoscedastic residuals from one step ahead forecasting $r_t:= (Y_t-\hat{\mu}_{t|t-1})/\hat{\sigma}(\bm{X}_t)$, the conventional kernel density estimate of $g$ is: 
 \begin{equation} \label{est_e}
\hat{g}(\epsilon)=\frac{1}{T}\sum_{t=1}^{T}\mathcal{K}(\epsilon-r_{t};h_r).
\end{equation}
By equation \eqref{model_e}, the conditional density of $Y_t$ can be estimated by:
 \begin{equation} \label{model_f}
\hat{f}_{t|t-1}(y)= \sum_{t=1}^{T}\mathcal{K}\left(y_t-r_{t};h_r\right)/\left\{\hat{\sigma}(\bm{X}_{t})T\right\},
 \end{equation}
where $y_t=(y-\hat{\mu}_{t|t-1})/\hat{\sigma}(\bm{X}_t)$. 

In the end, to make forecasting at  instant $T+1$, we have the conditional mean and density estimator:
\begin{equation} 
\hat{\mu}_{T+1|T}=\hat{m}(\bm{X}_{T+1})+\hat{\sigma}(\bm{X}_{T+1})\sum_{\tau=1}^{p}\hat{a}_\tau\hat{u}_{T+1-\tau},
\end{equation}
\begin{equation} 
\hat{f}_{T+1|T}(y)= \frac{\sum_{t=1}^{T}\mathcal{K}\left(y_{T+1}-r_{t};h_r\right)}{\{\hat{\sigma}(\bm{X}_{T+1})T\}}.
\end{equation}
\subsection{Further Discussion on Implementation}
\subsubsection{Choice of Kernel}
We mainly follow the kernel choices in \cite{lee2015power}. Gaussian kernel is used for each non-circular variable, i.e., $\mathcal{K}(u;h)=(1/\sqrt{2\pi}h)\exp [-u^2/(2h^2)]$.
Here $u$ is the Euclidean distance between two points of interest, and $h$ is the bandwidth. Von Mises kernel is used for circular variables, such as the wind direction, in which case 0 degrees and 360 degrees are identical. The Von Mises kernel is defined as $\mathcal{K}(u;h)=\exp[h^{-2} \cos(u)]/(2\pi I_0(h^{-2}))$. Here $I_0(\cdot)$ is the modified Bessel function of order 0, $1/h^2$ is the concentration parameter. When dealing with multivariate variables, we choose the multiplicative
kernel as the base kernel, i.e., $\mathcal{K}(\|\bm{u}\|;\bm{h}):= \Pi_{j=1}^d \mathcal{K}(u_j;h_j)$. When the input dimension goes higher, an additive structure is imposed on the multiplicative kernel, which has been shown effective in alleviating ``curse of dimensionality". To be more specific, each multiplicative kernel is limited to be product kernels of at most three inputs and the multivariate kernel is an average of all the multiplicative kernels. 
\subsubsection{Choice of Bandwidth}
Bandwidth is a key parameter for kernel performance. Theoretically speaking, the best bandwidth is the multivariate cross-validation. Nevertheless, multivariate cross-validation is extremely computationally costly \cite{lee2015power, khorramdel2018fuzzy}. Therefore, substitution methods are usually adopted for practical considerations, especially for the short-term modeling of this work, which may require updating of the bandwidth in real time. In this work, we choose heuristic bandwidth selectors as follows. 

Firstly, we use the direct plug-in methods to determine bandwidths for every variable separately. Direct plug-in kind of methods choose bandwidths by minimizing the mean integrated square error (MISE) or its asymptotic approximation. For mean function $\tilde{m}_j(x ; h_j)=\mathbb E[Y|X_j]$, its MISE is: 
\begin{equation}\label{mise1}
{\operatorname{\small{MISE}}} =\mathbb E\left[\int\{\tilde{m}_j(x; h_j)-m_j(x)\}^{2} f_j(x) dx \right].
\end{equation}
As a result, the direct plug-in bandwidths proposed by \cite{ruppert1995effective} is adopted for local linear regression in this work. Similarly, direct plug-in bandwidths are chosen for estimating $\sigma(\cdot)$. 

For density estimation, the MISE is defined as: 
\begin{equation}\label{mise2}
\operatorname{MISE}\{\tilde{g}(r ; h_r)\} =\mathbb{E} \int\left[\tilde{g}(r;h_r)-g(r)\right]^{2} d r,
\end{equation}
and $h_r$ can be estimated by the direct plug-in method proposed by \cite{sheather1991reliable}. Nevertheless, we further adjust $h_r$ to be adaptive to improve the estimation performance. The adaptive approach allows the bandwidth to vary for different observations and thus provides a calibration for the individual bandwidths and improves estimation accuracy at the tail of the densities where the data becomes sparser. Following \cite{silverman1986density}, the adaptive bandwidth can be
calculated as $h_{r,i}=h_r\{\bar{g}(r_i)/s\}^{-1/2},$ where
$s =\exp[n^{-1}\sum_{i=1}^T \log \bar{g}(r_i)]$, and
$\bar{g}(\cdot)$ is the conditional density function estimated using
bandwidth $h_r$.

We call the proposed bandwidth selector as univariate direct plug-in (U-DPI). In Table \ref{bw}, we compared U-DPI with univariate cross validation (U-CV) \cite{fan1996local}, and multivariate direct plug-in (M-DPI) \cite{wand1994multivariate} using our case study datasets. As can be seen, the proposed bandwidth selection works well. It performs better than M-DPI and computes much faster than U-CV with comparable accuracy. As a greedy procedure, we are also aware that the proposed bandwidth cannot be guaranteed to be the optimal selection. Nevertheless, as observed in the performance evaluation, DEAR with the chosen bandwidth is able to provide a remarkable performance improvement compared to the current state-of-art methods. 
\begin{table}
	\centering
	\caption{RMSE comparison for bandwidth selection}\label{bw}
	\begin{tabular}{c|ccccc}
		\toprule
		case&solar& WT1&WT2&WT3&WT4\\	
		\hline
		U-DPI&10.12&2.88&2.53&2.26&2.86\\	
		U-CV&11.36&3.49&3.11&2.37&2.77\\
		M-DPI&10.98&4.16&3.90&2.65& 2.97\\
		\bottomrule
	\end{tabular}	
\end{table}

\begin{algorithm}[t]
	\caption{DEAR's algorithm}   
	\label{alg}  
	\begin{algorithmic}[1]
		\REQUIRE $\mathcal{F}_{T}=\{(\bm{X}_t,Y_t),t=1,\cdots,T\}$
		\renewcommand{\algorithmicensure}{\textbf{Inference Procedure:}}   
		\ENSURE ~~ 
		\STATE Calculate the initial conditional mean and standard deviation estimator $\tilde{m}(\bm{X}_t)$ and $\tilde{\sigma}(\bm{X}_t)$ using $\mathcal{F}_{T}$.
		
		\STATE Initialize $\hat{m}(\bm{X}_t)=\tilde{m}(\bm{X}_t)$, $\hat{\sigma}(\bm{X}_t)=\tilde{\sigma}(\bm{X}_t)$. Obtain $\hat{u}_t$ and the autocorrelation order $p$.
		\STATE While the termination condition is not satisfied, do: 
		\begin{enumerate}[(1)]		
			\item Obtain $\hat{a}_1, \cdots,\hat{a}_p$ using $\hat{u}_t$.
			\item Calibrate $Y_t$ to $\tilde{Y}_t$ and obtain $\mathcal{\tilde{F}}_{T}=\{(\bm{X}_t,\tilde{Y}_t),t=1,\cdots,T\}$.			
			\item Update $\hat{m}(\bm{X}_t)$ and $\hat{\sigma}(\bm{X}_{t})$ using $\mathcal{\tilde{F}}_{T}$.
			\item Re-calculate $\hat{u}_{t}$ and $r_{t}$. Terminate the iteration if the Box-Ljung test on $r_{t}$ for all the $p$ lags are not rejected.
		\end{enumerate} 
		\STATE Record $\mathcal{R}_T=\{r_{t}\}$, and $\hat{\mathcal{F}}_T=\{(\bm{X}_t,\hat{Y}_t)\}$, where $\hat{Y}_t=Y_t-\hat\sigma(\bm{X}_{t})\sum_{\tau=1}^{p}\hat{a}_\tau\hat u_{t-\tau}$ and $t=1,\cdots T$.
		
		\STATE Calculate the bandwidth for density estimation $h_{r,i}$.
		\renewcommand{\algorithmicensure}{ \textbf{Estimation Procedure:}} 
		\ENSURE ~~  
		\STATE For the next instant $T+1$:	
		\begin{enumerate} [(1)]
			\item Estimate the conditional mean $\hat{\mu}_{T+1|T}$ and conditional density $\hat{f}(y|\bm{X}_{T+1},\mathcal F_T)$ using $\mathcal{\hat{F}}_{T}$.
			\item Update $\mathcal{\hat{F}}_{T}$ and $\mathcal{R}_{T}$ using $\hat{Y}_{T+1}$ and $r_{T+1}$.
		\end{enumerate} 
	\end{algorithmic}
\end{algorithm} 
\subsubsection{Implementation}
In practice, the power output is usually restrained in a reasonable region. As a result, we set the upper and lower limits on the predicted values to make them sensible. Besides, we check data sparsity by thresholding on the input density and substitute local linear regression with conventional kernel for conditional mean estimation, as recommended by \cite{taylor2013challenging}. Thirdly, in a long time span, the power curve can change due to degradation and maintenance. We therefore adopt a rolling window approach to update the power curve, i.e., to use the latest observations before the testing sample for estimation. The rolling window should contain enough observations to avoid curse of dimensionality. Based on the above explanations, the detailed algorithm for DEAR is summarized in Algorithm~\ref{alg}.

\section{Case Studies}
Two case studies are conducted to compare DEAR's performance with existing methods. For conditional density estimator, we choose additive multiplicative kernel (AMK) \cite{lee2015power}. For conditional mean estimator, we additionally compare DEAR with the initial estimate of conditional mean ($\tilde{m}(\cdot)$), i.e., additive multiplicative local linear regression (AML). Besides, Support Vector Regression (SVR) and Deep Belief Network (DBN) are included, which have been indicated to be competitive for renewable energy modeling \cite{ferlito2017comparative,wang2019review}. Last but not least, the persistent model is also added as is it hard to be outperformed for short lead times. In all case studies, the input variables are selected based on the RMSE value of the validation dataset. For SVR and DBN, we include both the weather conditions and lagged power generations as the input variables. 

Performance is evaluated by the following metrics. Firstly, the root mean square error (RMSE) is adopted for conditional mean
estimation,
\begin{equation}
	\textnormal{RMSE}\ = \sqrt{\frac{1}{N'}\sum_{t=1}^{N'}(\hat{\mu}_{t|t-1}-Y_t)^2},
\end{equation}
 where $N'$ is the size of testing dataset. Secondly, the mean continuous ranked probability score (CRPS) is adopted for the conditional density estimation,
 \begin{equation}
 	\textnormal{CRPS}= \frac{1}{N'}\sum_{t=1}^{N'}\int(\hat{F}(y|\bm{X}_t,\mathcal{F}_{t-1})-\mathbbm{1}(y>Y_t))^2 \ dy,
 \end{equation}
where $\hat{F}(y|\bm{X}_t,\mathcal{F}_{t-1})$ is the estimated conditional CDF and $\mathbbm{1}(\cdot)$ is the indicator function. 
Smaller CRPS indicates better estimations. Furthermore, to provide more comprehensive evaluation on the conditional density estimation, we consider the coverage probability deviation (Dev) \cite{Golestaneh2016} and predictive interval normalized average width (PINAW) \cite{zhang2014advanced} to evaluate the reliability and sharpness, respectively. Given the estimated upper quantile $\hat{U}_{ij}$ and lower quantile $\hat{L}_{ij}$ for $Y_i$ corresponding to the nominal quantile level $U_{j}$ and $L_j$ , the metrics are defined as 
\begin{equation}
\text{Dev}_j =|(U_j-L_j)-\frac{1}{N} \sum_{i=1}^{N} \xi_i|,
\end{equation}
where $\xi_i=1$ if $Y_i \in[\hat{L}_{ij},\hat{U}_{ij}]$, and $\xi_i=0$ otherwise, and
\begin{equation}
\text{PINAW}_j=\frac{1}{N R} \sum_{i=1}^{N} (U_{ij}-L_{ij}),
\end{equation}
where $R$ is the range of the underlying targets. As we focus on full density estimation in this work, we take an average over 38 quantiles with nominal levels from 0.025 to 0.975, at increments of 0.025, except for the median (0.5). This leading to the averaged Dev and averaged PINAW,
\begin{equation}
\text{aDev} =\frac{1}{M} \sum_{j=1}^{M}|(U_j-L_j)-\frac{1}{N} \sum_{i=1}^{N} \xi_i|,
\end{equation}
\begin{equation}
\text{aPINAW}=\frac{1}{MNR} \sum_{j=1}^{M}  \sum_{i=1}^{N} (U_{ij}-L_{ij}),
\end{equation}

\subsection{Datasets}
The solar power curve dataset is from the Global Energy Forecasting Competition 2014
(GEFCom2014) \cite{Hong2016}. The aim is to estimate the distribution
of solar power production given the weather variables. There are over
59000 hourly average records on solar power production and 12 weather
condition variables. Besides the weather condition, ``time of the day"
is also considered as an input variable due to its importance. The significant input variables are chosen to be ``time of the day" and "solar radiance". Moreover, evaluation based on RMSE has shown that ``time of
the day" is the most influential factor that affects conditional
standard deviation. The experiments use the 30001th to 32000th points as the testing dataset to conduct hourly ahead estimation.

The wind power curve dataset is from the wind farm “La Haute Borne” (Meuse, France), provided by ENGIE Renewable Energy\footnote{The dataset is available at https://opendata-renewables.engie.com/pages/home/}. The dataset contains 10-min average wind turbine data of 4 turbines (denoted as WT1-WT4) since 2013. Rough filtering is firstly conducted to filter out the idle/controlled periods. For wind power curve, wind speed and wind direction are always included for every three-dimensional kernel, as recommended by \cite{lee2015power}. The significant input variables are chosen to be  ``wind speed", ``wind direction", ``temperature" and ``turbulence intensity". Similar as the solar power case, wind speed is the most influential factor that affects conditional standard deviation. The experiments use the 36500th to 38500th points of year 2014 as the testing dataset to conduct 10-min ahead estimation. 

\subsection{Results}
The results of the solar and wind power curve estimation case studies are shown in in Table~\ref{rmse}. DEAR gets competitive conditional mean estimation regarding RMSE. DEAR significantly outperforms AML and AMK by considering temporal dependence. Besides, DEAR outperforms persistent model because it accounts for the weather variables, which are crucial for power curve. Ignoring weather conditions causes the unsatisfying performance of persistent model. Last but not least, It can be seen that SVR and DBN provide competitive results compared to other candidates, especially in the solar case, where the data contains higher uncertainty and the temporal dependence plays a more important role. Nevertheless, they do not have dominant advantage in terms of accuracy. One possible reason is that the number of covariates is small, and the relation between weather variables and power output is not very complex. As a result, the capacity of machine learning algorithms are not fully utilized, especially with limited datasets. Furthermore, in the wind power datasets, where the data has relatively low uncertainty, kernel based methods have more apparent advantages, as they effectively utilized the input variables and obtain stronger explain ability. More importantly, the superiority of DEAR lies in its conditional density estimation. As can be seen in Table \ref{density}, DEAR improves both the reliability and sharpness of the conditional density estimator, leading to a smaller CRPS value. 
	\begin{table}[t]
	\centering
	\caption{Performance comparison for the conditional mean estimation}\label{rmse}
	\begin{tabularx}{\linewidth}{p{0.6cm}|XXXXXX}
		\toprule
		Case&\multicolumn{6}{c}{RMSE}\\
		\hline
		&DEAR&AML&AMK&SVR&DBN&PER\\	
		solar&7.15&10.07&10.11&6.75&8.09&10.29\\	
		WT1&2.09&2.9&2.92&3.04&4.46&4.79\\
		WT2&2.22&2.51& 2.53 &3.78&6.34& 7.04\\										
		WT3 &2.03& 2.28& 2.29 &2.51&6.25& 7.21 \\	
		WT4&2.38&2.83&2.84&3.19&4.41&5.88\\		
		\bottomrule
	\end{tabularx}	
\end{table}
\begin{table}[t]
	\centering
	\caption{Performance comparison for the conditional density estimation}\label{density}
	\begin{tabularx}{\linewidth}{p{0.6cm}|p{0.8cm}p{0.8cm}|p{0.8cm}p{0.8cm}|XX}
		\toprule
		Case&\multicolumn{2}{c|}{aDev}&	\multicolumn{2}{c|}{aPINAW}&\multicolumn{2}{c}{CRPS}\\	
		& DEAR&AMK&DEAR&AMK&DEAR&AMK\\
		\hline		
		solar&0.0739&0.226&0.0351&0.0965&2.12&3.16\\		
		WT1&0.0161&0.103&0.0283&0.0518&1.06&1.24\\
		WT2&0.0240&0.0904&0.0331&0.0464&1.13&1.77\\
		WT3&0.0107&0.119&0.0314&0.0474&1.02&1.23\\
		WT4&0.0247&0.0839&0.0393&0.0608&1.3&1.63\\	
		\bottomrule
	\end{tabularx}	
\end{table}

To emphasize DEAR's effectiveness on short-term temporal dependence, we compare DEAR with KDES \cite{bessa2012time} in Table \ref{time} using the solar power case as an illustration. At instant $T$, KDES assigns forgetting factor $\lambda^{T-i}$ to the $i$th historical observation. KDES outperforms AML when a proper forgetting factor is used. Nevertheless, although smaller $\lambda$ is supposed to better capture short-term temporal dependence, it increases the RMSE, since smaller $\lambda$ causes computation problems for kernel because of sparsity. Overall, KDES does not outperform DEAR. KDES is more suitable for long-term temporal dependence, such as system degradation, whereas DEAR outperforms KDES on short-term modeling.
\begin{table}[t]
	\centering
	\caption{RMSE of DEAR, AML and KDES. }\label{time}
	\begin{tabularx}{\linewidth}{p{0.7cm}p{0.7cm}|XXX}
		\toprule
		DEAR&AML&KDES ($\lambda$=0.999)&KDES ($\lambda$=0.995)&KDES ($\lambda$=0.95)\\ 
		\hline
	    7.15&10.07&9.94&10.65&16.77\\
		\bottomrule
	\end{tabularx}	
\end{table}

In the following, we use plots to provide more intuitions. 
\begin{figure}[t]
	\centering  
	\subfloat[Conditional mean of AMK]{\includegraphics[width=0.48\linewidth]{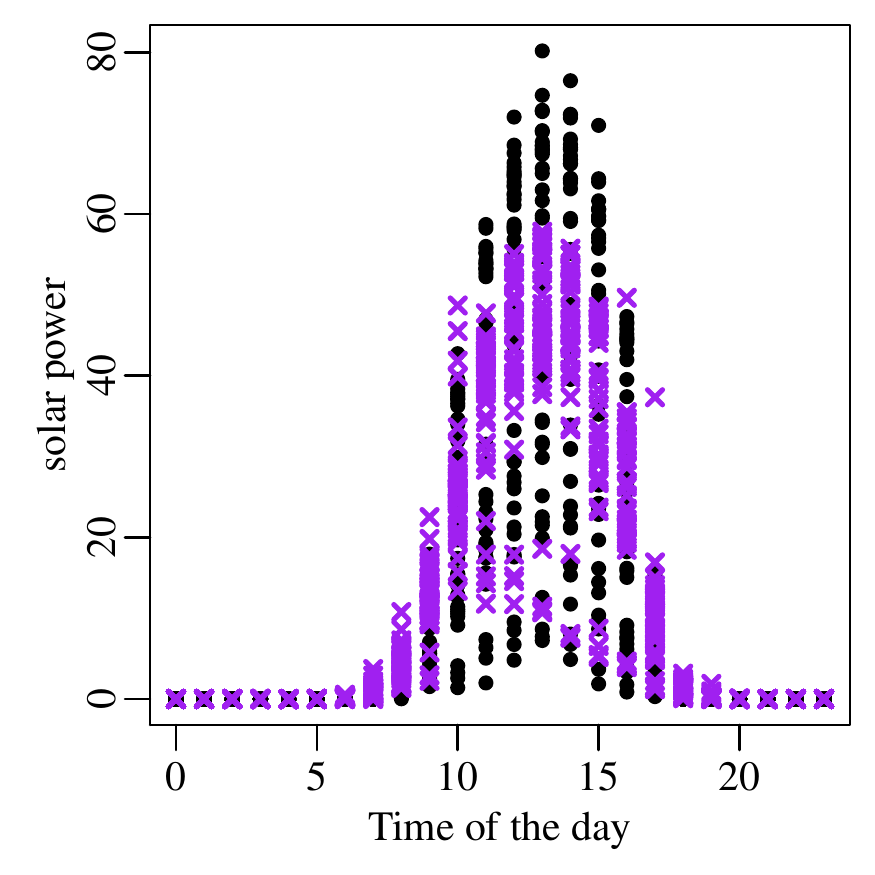}\label{mean_1}}
	\subfloat[Conditional mean of DEAR]{\includegraphics[width=0.48\linewidth]{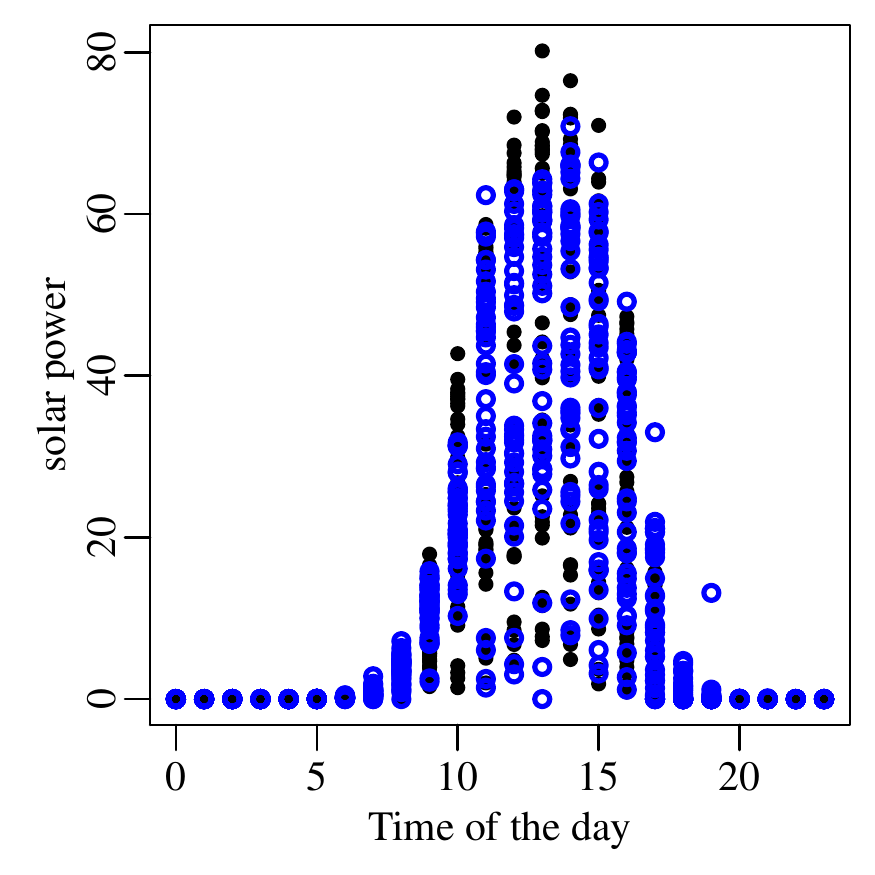}\label{mean_2}}
	\caption{Conditional mean estimation visualization; Case: solar power curve.}\label{mean_illustration}
	\subfloat[Residual confidence interval]{	\includegraphics[width=0.48\linewidth]{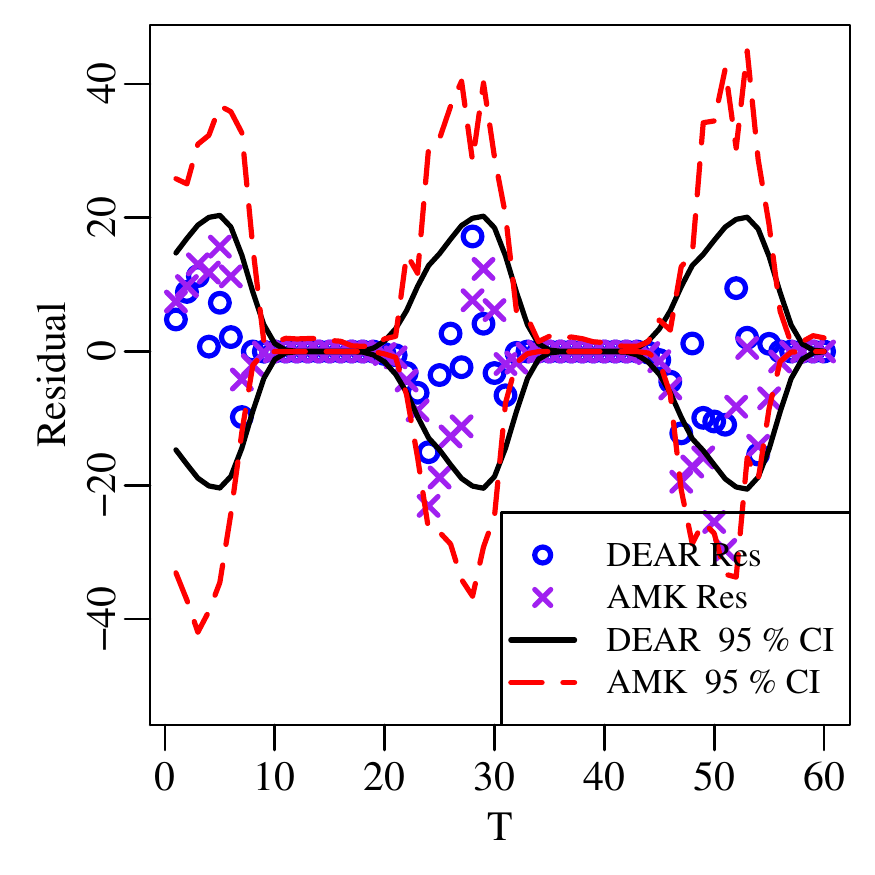}\label{ci_1}}
	\subfloat[Power confidence interval]{\includegraphics[width=0.48\linewidth]{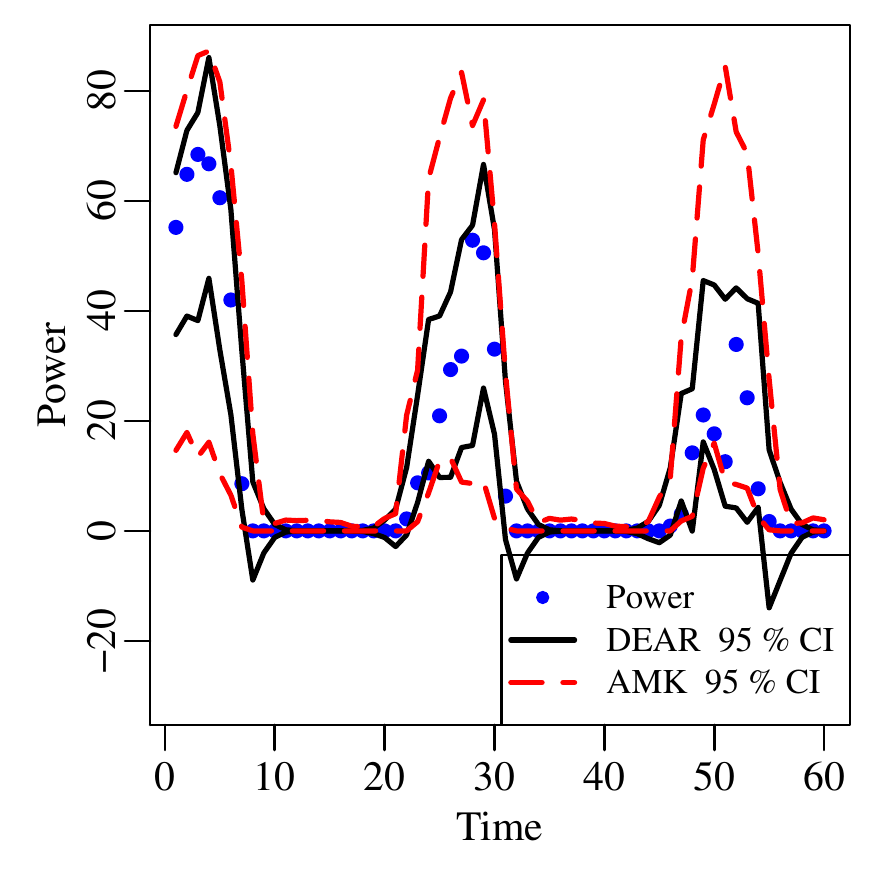}\label{ci_2}}
	
	\subfloat[Residual confidence interval]{\includegraphics[width=0.48\linewidth]{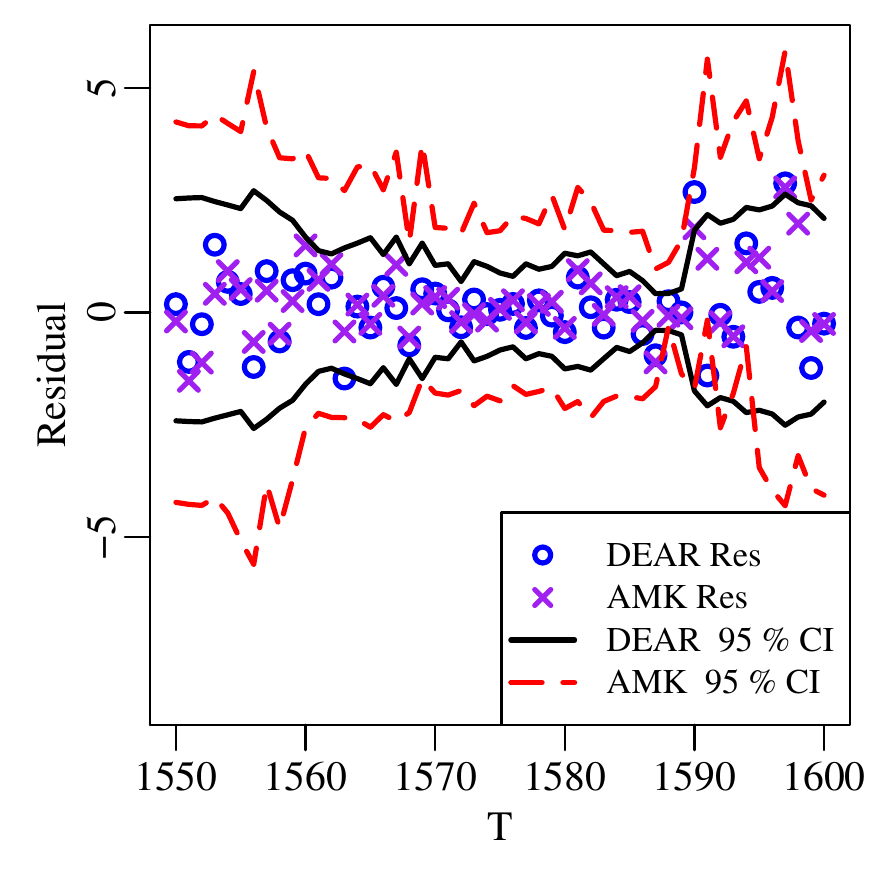}\label{ci_3}}
	\subfloat[Power confidence interval]{\includegraphics[width=0.48\linewidth]{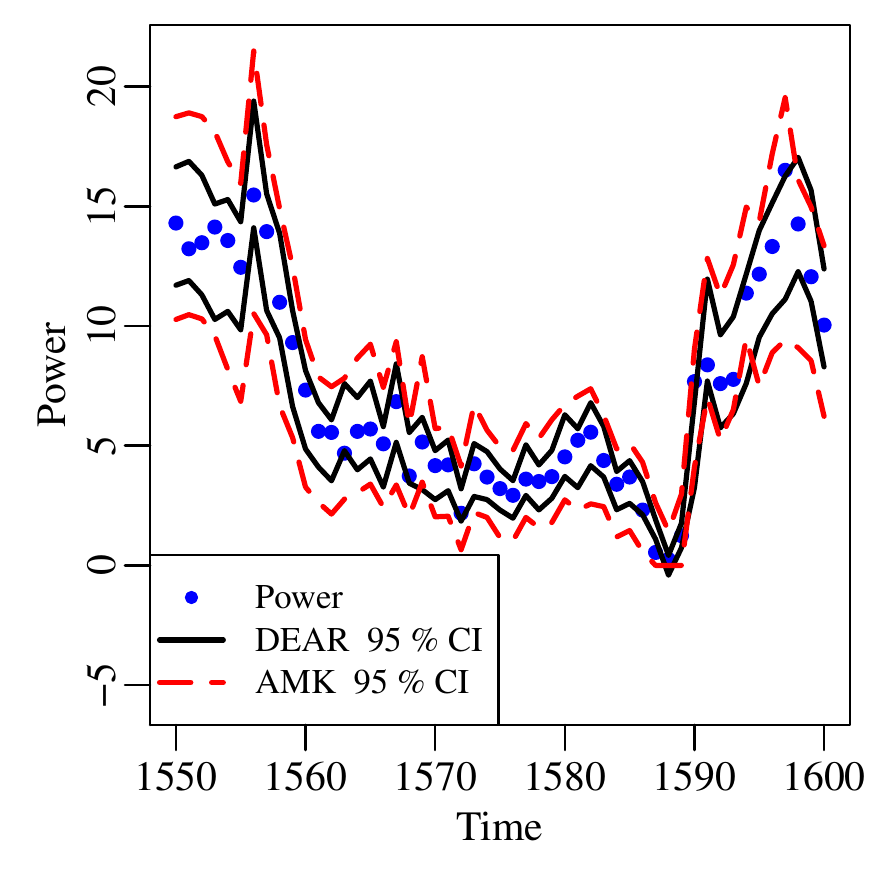}\label{ci_4}}
	\caption{Conditional density estimation visualization; Figure (a)-(b): solar power curve; Figure (c)-(d): wind power curve} \label{density_intuition}
\end{figure}
Fig. \ref{mean_1}-\ref{mean_2} plot the conditional mean
estimation. DEAR performs better especially when the observations have
larger deviations from theoretical power curve. Fig. \ref{ci_1}-\ref{ci_4} plot the conditional density
estimation. In Fig. \ref{ci_1} and \ref{ci_3}, DEAR has both smaller residual and
confidence interval compared to AMK. The CRPS and aPINAW value verify that this
smaller confidence interval is better. In Fig. \ref{ci_2}, the
confidence intervals against the real values of power ($Y$) are also
shown. It can be seen that the confidence interval of DEAR moves
better with the real values of $Y$, in a sense that it makes
adjustments in response to the short-term data features. Moreover,
since DEAR accounts for the autocorrelation, even with the same weather
inputs, the mean prediction and its confidence interval can be
different at different instants. Overall, Fig. \ref{mean_illustration}-\ref{density_intuition} indicate that DEAR improves estimation on both conditional mean and density estimation.

\section{Conclusion}
Renewable power generation has grown rapidly and become indispensable for energy security and global warming mitigation. Since the renewable energy power curve are hard to model by first principles, accurate modeling using data driven methods is essential for RE's control strategies, security of electrical grid and power trading etc. In this paper, we propose DEAR to account for temporal dependence when conducting conditional density estimation, which effectively improves both conditional mean and density estimators. Real-field case studies from wind and solar energy are conducted for showcase. In the future, further explorations can be done on potentially nonlinear dependency between power series and its lagged series as well as considering spatial effect (intra-system interactions) for larger scale systems. Furthermore, the proposed estimators can be used for further applications such as real time condition monitoring for RE equipments. 

\setlength{\lineskip}{0.1em}
\small
\bibliographystyle{IEEEtran}
\bibliography{powercurve_paper}	
\vspace*{-4\baselineskip}
\begin{IEEEbiographynophoto}{Yuchen SHI}
received the B.S degree in industrial engineering from Nanjing University, China, in 2016. She is currently
a Ph.D. candidate in the Department of Industrial
Systems Engineering and Management, National University of Singapore, and concurrently a doctoral researcher at the Future Resilient Systems programme, Singapore-ETH Centre.
Her research interests include statistical learning and condition monitoring.
\end{IEEEbiographynophoto}
\vspace*{-4\baselineskip}
\begin{IEEEbiographynophoto}{Nan CHEN}
	Nan Chen received the B.S. degree in automation
from Tsinghua University, Beijing, China, in 2006,
the M.S. degree in computer science in 2009, and
the M.S. degree in statistics and the Ph.D. degree
in industrial engineering from the University of
Wisconsin-Madison, Madison, WI, USA, both in
2010. He is currently an Associate Professor with
the Department of Industrial Systems Engineering
and Management, National University of Singapore,
Singapore. His research interests include statistical
modeling and surveillance of engineering systems,
simulation modeling design, condition monitoring, and degradation modeling.
\end{IEEEbiographynophoto}

\end{document}